\documentclass[aps,prd,twocolumn,superscriptaddress,amsmath,amssymb]{revtex4}
\usepackage{graphicx} 
\usepackage{epstopdf} 
\usepackage{dcolumn} 
\usepackage{xcolor} 
\usepackage{amsmath,amssymb} 
\usepackage[hidelinks]{hyperref} 
\usepackage[utf8]{inputenc} 
\usepackage[english]{babel} 
\usepackage[displaymath, mathlines]{lineno} 
\usepackage{blindtext} 
\usepackage{ifthen} 
\usepackage{orcidlink} 
\usepackage{multirow}
\hypersetup{
    breaklinks=true,   
}

\newboolean{articletitles}
\setboolean{articletitles}{true} 

\usepackage{environ}
\NewEnviron{comment}{}
\newcommand\con{\RenewEnviron{comment}{\color{olive}\BODY}}

\con
\input{belle2-symbols}

\begin{document}

\title{Observation of time-dependent \CP violation and measurement of the branching fraction of \jpsipiz decays}

\author{I.~Adachi\,\orcidlink{0000-0003-2287-0173}} 
  \author{L.~Aggarwal\,\orcidlink{0000-0002-0909-7537}} 
  \author{H.~Ahmed\,\orcidlink{0000-0003-3976-7498}} 
  \author{H.~Aihara\,\orcidlink{0000-0002-1907-5964}} 
  \author{N.~Akopov\,\orcidlink{0000-0002-4425-2096}} 
  \author{A.~Aloisio\,\orcidlink{0000-0002-3883-6693}} 
  \author{N.~Althubiti\,\orcidlink{0000-0003-1513-0409}} 
  \author{N.~Anh~Ky\,\orcidlink{0000-0003-0471-197X}} 
  \author{D.~M.~Asner\,\orcidlink{0000-0002-1586-5790}} 
  \author{H.~Atmacan\,\orcidlink{0000-0003-2435-501X}} 
% ...
\collaboration{The Belle II Collaboration}

\begin{abstract}
We present a measurement of the branching fraction and time-dependent charge-parity (\CP) decay-rate asymmetries in \jpsipiz decays.
The data sample was collected with the \belletwo detector at the SuperKEKB asymmetric \epem collider in 2019--2022 and contains \nbb \BBbar meson pairs from \FourS decays. 
We reconstruct \njpsipiz signal decays and fit the \CP parameters from the distribution of the proper-decay-time difference of the two \B mesons.
We measure the branching fraction to be $\bf(\jpsipiz)=(\bfval \pm \bferr \pm \bfsys)\times 10^{-5}$ and the direct and mixing-induced \CP asymmetries to be $\ccp=\ccpval \pm \ccperr \pm \ccpsys$ and $\scp=\scpval \pm \scperr \pm \scpsys$, respectively, where the first uncertainties are statistical and the second are systematic.
We observe mixing-induced \CP violation with a significance of $5.0$ standard deviations for the first time in this mode.
\end{abstract}

\maketitle

\section{Introduction}
\label{sec:introduction}
Precision measurements of \CP asymmetries are powerful experimental tools to indirectly probe physics beyond the Standard Model (SM).
In the SM, \CP violation is governed by a single complex phase in the Cabibbo-Kobayashi-Maskawa (CKM) quark mixing matrix~\cite{PhysRevLett.10.531, Kobayashi:1973fv}. 
The unitarity of the CKM matrix can be represented as a triangle in the complex plane.
Precise measurements of the angle $\phi_1=\arg(-\Vcd\Vcb^*/\Vtd\Vtb^*)$~\cite{convention}, where $V_{ij}$ are the CKM matrix elements, have been performed in \jpsikz decays~\cite{PhysRevD.79.072009, PhysRevLett.108.171802, PhysRevLett.132.021801}.
These decays proceed via tree-level \ccs transitions.
However, decay amplitudes involving the emission and reabsorption of a \W boson, also known as ``penguin'' diagrams, occur at higher order in SM perturbation theory and can induce a shift in the measurement of $\phi_1$, thereby limiting the sensitivity of CKM fits~\cite{GROSSMAN1997241}.
In the absence of penguin amplitudes, the direct and mixing-induced \CP asymmetries are predicted to be $C_{\jpsi \Kz}=0$ and $-\eta S_{\jpsi \Kz}=\sin2\phi_1$, where $\eta$ is the \CP eigenvalue of the decay final state.
The world-average values, $C_{\jpsi \Kz}=0.009\pm0.010$ and $-\eta S_{\jpsi \Kz}=0.708\pm0.012$~\cite{HeavyFlavorAveragingGroup:2022wzx}, are in agreement with independent constraints on the CKM matrix~\cite{Charles:2004jd,UTfit:2005ras}.

The decay \jpsipiz proceeds via color-suppressed \ccd tree-level transitions and its \CP asymmetries can be used to constrain the contributions from penguin topologies in \jpsikz.
In the presence of penguin amplitudes, the observable phase measured by the mixing-induced \CP asymmetry is $\phi_d^{\rm eff}=\phi_d+\Delta \phi_d$, where $\Delta \phi_d$ is a shift of the order of $0.5^\circ$ from the SM value of $\phi_d = 2\phi_1$~\cite{Ciuchini:2005mg, Barel_2021}.
In addition, the value of the branching fraction is used to probe the size of non-factorizable $SU(3)$-breaking effects, which are the main contributions to the theoretical uncertainties in the extraction of $\Delta \phi_d$~\cite{Barel:2023oa}.
The world-average value of the branching fraction is based on the measurements from the \babar~\cite{BaBar:2008kfx}, \belle~\cite{Belle:2018nxw}, and CLEO~\cite{PhysRevD.62.051101} experiments, resulting in $\BF(\jpsipiz)=(1.66\pm 0.10)\times 10^{-5}$~\cite{PhysRevD.110.030001}.
More recently, \lhcb measured the ratio of branching fractions of \jpsipiz and \jpsikstp decays~\cite{LHCb:2024ier}, resulting in a comparable precision on $\BF(\jpsipiz)$ using the current knowledge of $\BF(\jpsikstp)$.
\babar~\cite{BaBar:2008kfx} and \belle~\cite{Belle:2018nxw} also measured the direct and mixing induced \CP asymmetries.
The world average values are $C_{\jpsipiz}=0.04\pm0.12$ and $S_{\jpsipiz}=-0.86\pm0.14$~\cite{HeavyFlavorAveragingGroup:2022wzx}.

The current values of $\phi_d$ and $\Delta \phi_d$, based on the analysis in Ref.~\cite{Barel_2021}, are $(44.4^{+1.6}_{-1.5})^\circ$ and $(-0.73^{+0.60}_{-0.91})^\circ$, respectively, and do not yet include the most recent measurements from \lhcb~\cite{PhysRevLett.132.021801,LHCb:2024ier}.
With an improvement by a factor of two on the experimental precision on \jpsikz only, the precision on $\phi_d$ would be limited to $1^\circ$ by the uncertainty on $\Delta \phi_d$.
On the other hand, a similar improvement on the precision of both \jpsikz and \jpsipiz would improve the precision on $\phi_d$ to $0.78^\circ$ and confirm the presence of nonzero penguin contributions~\cite{Barel_2021}.
Therefore, the current experimental knowledge on \jpsipiz should be improved.

Here we present a measurement of the branching fraction and \CP asymmetries in \jpsipiz decays using a sample of energy-asymmetric \epem collisions at the \FourS resonance provided by the SuperKEKB accelerator~\cite{Akai:2018mbz} and collected with the \belletwo detector~\cite{Abe:2010gxa}.
The sample corresponds to \lumi of integrated luminosity and contains \nbb \BBbar events~\cite{Belle-II:2024vuc}.
An additional \offres off-resonance sample recorded at $60$~\mev below the \FourS is used to model background from continuum $\epem \to \qqbar$ events, where $\qqbar$ indicates pairs of \uquark, \dquark, \squark, or \cquark quarks.

The \CP asymmetries are determined from the distribution of the proper-decay-time difference of $\Bz\Bzb$ pairs.
We denote pairs of \Bz mesons as \bsig and \btag, where \bsig decays into a \CP-eigenstate at time \tcp, and \btag decays into a flavor-specific final state at time \ttag.
For quantum-correlated neutral \B-meson pairs from \FourS decays, the flavor of \bsig is opposite to that of \btag at the instant when the first \B decays.
The probability to observe a \btag meson with flavor $q$ ($q=+1$ for \Bz and $q=-1$ for \Bzb) and a proper-time difference $\Delta t \equiv \tcp - \ttag$ between the \bsig and \btag decays is
\begin{linenomath}
\begin{align}
\begin{split}
\mathcal{P}(\dt,q)= \frac{e^{-|\dt|/\taud}}{4\taud}  \Big\{ 1 + q\big[ &  \scp \sin(\dmd \dt)\\
    - & \ccp \cos (\dmd \dt) \big] \Big\},
\end{split}
\label{eq:dt_theo}
\end{align}
\end{linenomath}
where \taud is the \Bz lifetime and \dmd is the mass difference between the $\Bz$ mass eigenstates~\cite{PhysRevD.110.030001}.

We fully reconstruct \bsig in the $\jpsi \piz$ final state using the intermediate decays \jpsill (with $\ell^{\pm}$ being an electron or a muon) and $\piz\to\gamma\gamma$, while we only determine the decay vertex of the \btag decay.
The flavor of the \btag meson is inferred from the properties of all charged particles in the event not belonging to \bsig~\cite{PhysRevD.110.012001}.
We first extract the signal yields from the distributions of the signal \bsig candidates in observables that discriminate against backgrounds, and then fit the \CP asymmetries from the \dt distribution of candidates populating the signal-enriched region.
We validate our analysis and correct for differences between data and simulation using \jpsikstp and \jpsiks decays, which are ten-fold more abundant than the expected signal and have a similar final state.
To reduce experimental bias, the signal region in data is examined only after the entire analysis procedure is finalized.
Charge-conjugated modes are included throughout the text.

The paper is organized as follows.
In Sec.~\ref{sec:experiment} we describe the experimental setup and in Sec.~\ref{sec:selection} we describe the reconstruction of signal candidates and the selection used to suppress the backgrounds.
The signal extraction and \CP asymmetry fits, from which the physics observables are measured, are detailed in Sec.~\ref{sec:de-fit} and Sec.~\ref{sec:dt-fit}, respectively.
The sources of systematic uncertainties are discussed in Sec.~\ref{sec:systematics}.
Finally, the results are summarized in Sec.~\ref{sec:summary}.
\section{Experiment}
\label{sec:experiment}
The \belletwo detector operates at the SuperKEKB accelerator at KEK, which collides 7~\gev electrons with 4~\gev positrons.
The detector is designed to reconstruct the decays of heavy-flavor hadrons and $\tau$ leptons.
It consists of several subsystems with a cylindrical geometry arranged around the interaction point (IP).
The innermost part of the detector is equipped with a two-layer silicon-pixel detector (PXD), surrounded by a four-layer double-sided silicon-strip detector (SVD)~\cite{Belle-IISVD:2022upf}.
Together, they provide information about charged-particle trajectories (tracks) and decay-vertex positions.
Of the outer PXD layer, only one-sixth is installed for the data used in this work.
The momenta and electric charges of charged particles are determined with a 56-layer central drift-chamber (CDC).
Charged-hadron identification (PID) is provided by a time-of-propagation counter and an aerogel ring-imaging Cherenkov counter, located in the central and forward regions outside the CDC, respectively.
The CDC provides additional PID information through the measurement of specific ionization.
Energy and timing of photons and electrons are measured by an electromagnetic calorimeter made of CsI(Tl) crystals, surrounding the PID detectors.
The polar angle coverage of the calorimeter is $12.4^\circ<\theta<31.4^\circ$, $32.2^\circ<\theta<128.7^\circ$ and $130.7^\circ<\theta<155.1^\circ$ in the forward, barrel and backward regions, respectively.
The tracking and PID subsystems, and the calorimeter, are surrounded by a superconducting solenoid, providing an axial magnetic field of 1.5~T.
The central axis of the solenoid defines the $z$ axis of the laboratory frame, pointing approximately in the direction of the electron beam.
Outside of the magnet lies the muon and \KL identification system, which consists of iron plates interspersed with resistive-plate chambers and plastic scintillators.

We use Monte Carlo simulated events to model signal and background distributions, study the detector response, and test the analysis procedure.
Quark-antiquark pairs from \epem collisions, and hadron decays, are simulated using \kkmc~\cite{Jadach:1999vf} with \pythia~\cite{Sjostrand:2014zea}, and \evtgen~\cite{Lange:2001uf} software packages, respectively.
The detector response is simulated using the \geant~\cite{Agostinelli:2002hh} software package.
The effects of beam-induced backgrounds are included in the simulation~\cite{Lewis:2018ayu, Liptak:2021tog}.
We use a simulated sample of generic \epem collisions, corresponding to a luminosity of approximately four times that of the experimental dataset.
We also use large samples of simulated \BBbar pairs, where one of the \B mesons is forced to decay into the final state of interest, while the other \B meson in the event is decayed inclusively.
One sample is used to study the signal, where the \B meson decays as \jpsipiz.
The other samples are used to study the dominant sources of backgrounds, where the \B meson decays inclusively into charmonium \jpsix modes.
Collision data and simulated samples are processed using the \belletwo analysis software~\cite{Kuhr:2018lps,basf2-zenodo}.
\section{Event selection}
\label{sec:selection}
Events containing a \BBbar pair are selected online by a trigger system based on the track multiplicity and total energy deposited in the calorimeter.
We reconstruct \jpsipiz decays using \jpsill and $\piz\to\gamma\gamma$ decays, in which the two light lepton tracks are reconstructed using information from the PXD, SVD, and CDC~\cite{Bertacchi:2020eez}. 
All tracks are required to have polar angles within the CDC acceptance ($17^\circ<\theta<150^\circ$).
Tracks used to form \jpsi candidates are required to have a distance of closest approach to the IP of less than $2.0$~\cm along the $z$ axis and less than $0.5$~\cm in the transverse plane to reduce contamination from tracks not generated in the collision.
Muons are identified using the discriminator $\prob_{\mu} = \lh_{\mu}/(\lh_{e}+\lh_{\mu}+\lh_{\pi}+\lh_{K}+\lh_{p}+\lh_{d})$, where the likelihood $\lh_{i}$ for each charged particle hypothesis combines particle identification information from all subdetectors except for the PXD and SVD.
Electron identification is provided by a boosted-decision-tree (BDT) classifier that combines several calorimeter variables and particle identification likelihoods~\cite{ebdt}.
We classify tracks as muons or electrons based on a loose PID requirement which is more than $95\%$ efficient on signal while rejecting more than $90\%$ of misidentified tracks.
The momenta of electrons are corrected for energy loss due to bremsstrahlung by adding the four-momenta of photons with energy in the lab-frame within $[75,1000]$~\mev and detected within 50~\mrad of the initial direction of the track.
The \jpsi candidates are formed by combining the four-momenta of oppositely charged lepton pairs having an invariant mass $\mll\in[2.9,3.2]$~\gevcc, where the average \jpsi mass resolution is approximately $13$~\mevcc for the muon mode and $16$~\mevcc for the electron mode.
Photons used to reconstruct \piz candidates are identified from calorimeter energy deposits greater than 22.5~\mev in the forward region and  20~\mev in the backward and barrel regions.
Photon energy corrections are derived from control samples reconstructed in collision data and applied to correct for the imperfect calorimeter energy calibration.
The \piz candidates are formed by combining pairs of photons with an invariant mass $\mgg\in[0.05,0.2]$~\gevcc, where the average \piz mass resolution is approximately $8$~\mevcc.

The beam-energy constrained mass \mbc and energy difference \deltae are computed for each \jpsipiz candidate as $\mbc \equiv \sqrt{(\ebeam/c^2)^2-(|\pstar|/c)^2}$ and $\deltae \equiv \estar - \ebeam$, where \ebeam is the beam energy, and \estar and \pstar are the reconstructed energy and momentum of the \bsig candidate, respectively, all calculated in the center-of-mass (c.m.)\ frame.
Signal \bsig candidates peak at the known \Bz mass~\cite{PhysRevD.110.030001} in \mbc and zero in \deltae.
The average \mbc and \deltae resolution for properly reconstructed \jpsipiz decays is approximately $5$~\mevcc and $50$~\mev, respectively.
Misreconstructed candidates from \BBbar events decaying into final states different than the signal peak in \mbc and follow an exponentially falling distribution in \deltae. 
Continuum events are uniformly distributed in \mbc and \deltae.
Only candidates satisfying $\mbc>5.2~\gevcc$ and $|\deltae|<0.5~\gev$ are retained for further analysis.

The \jpsipiz decay vertex is determined using the \treefit algorithm~\cite{HULSBERGEN2005566, Krohn:2019dlq}. 
The \bsig candidate is constrained to point back to the IP and the \jpsi and \piz masses are constrained to their known values~\cite{PhysRevD.110.030001}.
We retain only \bsig candidates with a successful vertex fit and $|\deltae|<0.3$~\gev.
The \btag decay vertex is reconstructed using the remaining tracks in the event.
Each track is required to have at least one measurement point in both the SVD and CDC subdetectors and correspond to a total momentum greater than 50~\mevc.
The \btag decay-vertex position is fitted using the \rave algorithm~\cite{Waltenberger_2008}, which allows for downweighting the contributions from tracks that are displaced from the \btag decay vertex, and thereby suppresses biases from secondary charm decays.
The decay-vertex position is determined by constraining the \btag direction, as determined from its decay vertex and the IP, to be collinear with its momentum vector~\cite{btube-conf}.
We only retain candidates with a successful tag-side vertex fit.
The proper-time difference between \bsig and \btag is estimated from the signed distance, $\Delta l$, of the \bsig and \btag decay-vertex positions along the \FourS boost direction,
\begin{equation}
\Delta\tilde{t} = \frac{\Delta l}{\betagamma\gamma^* c},
\label{eq:dt-approx}
\end{equation}
where $\betagamma=\boost$ is the \FourS Lorentz boost and $\gamma^*= 1.002$ is the Lorentz factor of the \B mesons in the c.m.\ frame.
We partially correct for the bias in $\Delta\tilde{t}$ arising from the angular distribution of the \B meson pairs in the c.m.\ frame ~\cite{Bevan:2014iga},
\begin{equation}
\deltat = \frac{\Delta\tilde{t} - (\beta^*/\beta) \cos\theta_{\rm c.m.}\tau_{B^0}}{1+s (\beta^* / \beta) \cos\theta_{\rm c.m.}},
\label{eq:dt-corr}
\end{equation}
where $\beta^*\approx0.06$ is the boost factor of the \Bz in the c.m.\ frame, $\theta_{\rm c.m.}$ is the polar angle of \bsig in the c.m.\ frame, and $s$ is the sign of $\Delta l$.
The residual bias in $\deltat$ and its impact on the \CP asymmetries is taken into account in the systematic uncertainties.
We retain candidates with $|\deltat|<10$~\ps and estimated uncertainty $\deltaterr\in[0.05,3]$~\ps.

We reduce the contribution from continuum events by requiring the zeroth to the second Fox-Wolfram moment~\cite{PhysRevLett.41.1581} to be less than $0.5$, which is more than $99\%$ efficient on signal while rejecting almost half of the continuum background.
We further reduce the continuum events by using a BDT classifier~\cite{ChenG16} combining several variables that discriminate between signal and background
The variables included in the BDT are the following, in decreasing order of discriminating power:
the cosine of the angle between the momentum of the positively charged lepton and the direction opposite to the momentum of the \Bz in the \jpsi frame; 
the ``cone'' variables developed by the CLEO collaboration~\cite{PhysRevD.53.1039};
the second to fourth harmonic moments calculated with respect to the thrust axis;
the ratio of the zeroth to the second and the zeroth to the fourth Fox-Wolfram moments; 
the modified Fox-Wolfram moments introduced in Ref.~\cite{PhysRevLett.91.261801};
the sphericity and aplanarity of the event~\cite{PhysRevD.1.1416};  
the cosine of the angle between the thrust axis of the \bsig and the thrust axis of the rest of the event;
and the event thrust~\cite{Brandt:1964sa, Farhi:1977sg}.
We impose a minimum threshold on the BDT output to maximize signal efficiency while rejecting as much background as possible.
This is achieved by choosing the threshold corresponding to the edge of the plateau of the signal efficiency \textit{vs.} continuum background rejection curve.
This requirement retains more than $97\%$ of the signal while rejecting more than $88\%$ of the remaining continuum background in simulation.

We further enrich the sample in signal by requiring at least one of the two leptons from the \jpsi candidate to fulfill a tight PID requirement.
We choose the requirement corresponding to the edge of the plateau of the signal efficiency \textit{vs.} misidentified lepton rejection curve.
This selection is more than $99\%$ efficient on signal while rejecting more than half of the misidentified tracks.
We suppress the contributions from beam backgrounds and misreconstructed energy deposits using a BDT classifier combining several calorimeter cluster variables~\cite{pbdt}.
In order to improve the resolution and reduce the correlation with \deltae, we redefine the beam-constrained mass as $\mbc^\prime$, by replacing the measured \piz momentum with $p^{*\prime}_{\piz}=\sqrt{(\ebeam-E^{*}_{\jpsi})^2/c^2-m^2_{\piz}c^2}\times\frac{p^{*}_{\piz}}{|p^{*}_{\piz}|}$, where $E^{*}_{\jpsi}$ is the energy of the \jpsi candidate in the c.m.\ frame, 
We only keep candidates with $\mbc^\prime>5.27~\gevcc$, $\deltae\in[-0.18,0.3]~\gev$, $\mll\in[2.95,3.2]~\gevcc$ and $\mgg\in[0.1,0.16]~\gevcc$.

Events with more than one candidate account for approximately $3\%$ of the data.
No candidates with different lepton mass hypotheses belonging to the same event are found in the data.
We keep the candidate with the reconstructed \mgg mass closest to the known \piz mass~\cite{PhysRevD.110.030001}.
This requirement selects the correct signal candidate more than $75\%$ of the time for events with multiple candidates in simulation. 

The same event selection is applied on the control channels, except for the reconstruction of the \KS candidate in \jpsiks and the requirements on the charged kaon track and invariant mass of the $\Kstarp\to \Kp \piz$ candidate in \jpsikstp. 
In order to reproduce the topology of \jpsipiz, we remove the additional tracks in the final state from the vertex fit of the control modes.

In the selection of the off-resonance sample, we remove the continuum suppression BDT and the following signal enhancement requirements, in order to increase the sample size. 
We verify in simulation that off-resonance and on-resonance continuum data passing the partial and full signal selections have similar distributions in the fit observables.
\section{Signal extraction fit}
\label{sec:de-fit}

The sample passing the event selection is populated by \jpsipiz candidates coming from signal events and backgrounds.
We classify as signal those candidates reconstructed from underlying \jpsipiz decays for which the \jpsi is properly reconstructed.
This includes a small contribution from candidates with a misreconstructed \piz, which accounts for approximately $3\%$ of the total signal yield.
Their distribution is centered around zero in \deltae and around the value of the \jpsi mass in \mll.
Among the sources of backgrounds, we classify as \jpsix those for which the \jpsi is properly reconstructed but originate from a different decay than the signal.
They follow an exponentially falling distribution in \deltae and have the same distribution in \mll as the signal.
In addition, we separate \BBbar events with a misreconstructed \jpsi and continuum backgrounds, both of which have a smooth distribution in \deltae and \mll.

We extract the signal yields from an extended likelihood fit to the unbinned \deltae and \mll distributions. 
The likelihood function is
\begin{equation}
\begin{split}
\lh = \frac{e^{-(\nsig+\nbkg)}}{N!}
\prod_{i=1}^{N} \Bigg \{
 \nsig P_{\jpsi \piz}^i +
\nbkg \Big[ \fbkg P_{\qqbar}^i \\
+ \frac{n_{\BBbar}}{\nbkg} P_{\BBbar}^i 
+ \frac{\nbkg(1-\fbkg)-n_{\BBbar}}{\nbkg} P_{\jpsi X}^i
\Big]
\Bigg \}
\end{split}
\label{eq:lh}
\end{equation}
where $i$ is the index of the candidate, $N$ is the total number of candidates in the dataset, \nsig and \nbkg are the signal and background yields, respectively, \fbkg is the fraction of continuum background, $n_{\BBbar}$ is the \BBbar background yield and $P^i=P(\deltae^i)P(\mll^i)$ is the probability distribution function (\pdf) of the $i$th candidate.

The \pdfs of the signal in the \deltae and \mll distributions are described by Crystal Ball functions~\cite{Gaiser:Phd,Skwarnicki:1986xj}.
The parameters of the \deltae \pdf are determined from simulation.
We account for differences between data and simulation by adjusting the mean and width according to differences observed for the \jpsikstp control sample. 
These adjustments consist of shifting the mean by \mush~\mev and scaling the width by a factor of \sgsf, where the uncertainties are statistical only.
The parameters of the \mll \pdf are determined separately for the \jpsiee and \jpsimm modes from a fit to the \jpsiks control sample in data.
The distribution of the \jpsix backgrounds is described by the sum of two exponential functions in \deltae and by a Crystal Ball function in \mll.
The parameters of the \deltae \pdf are determined from simulation while the parameters of the \mll \pdf are shared with the signal.
The \deltae and \mll distributions of the \BBbar backgrounds are described by exponential \pdfs with parameters determined from simulation.
The \deltae and \mll distributions of the continuum background are described by exponential \pdfs with parameters determined from the off-resonance data.

\begin{figure}[t]
\centering
\includegraphics[width=0.45\textwidth]{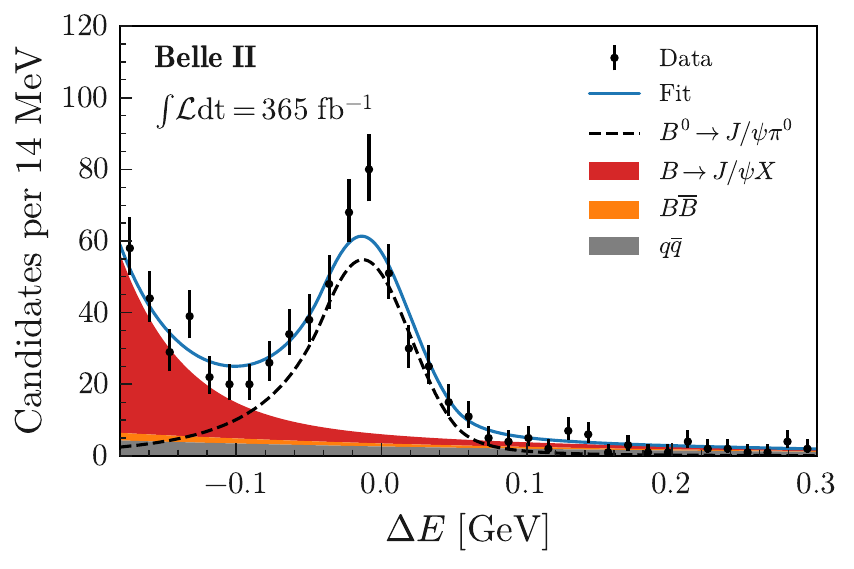}
\includegraphics[width=0.45\textwidth]{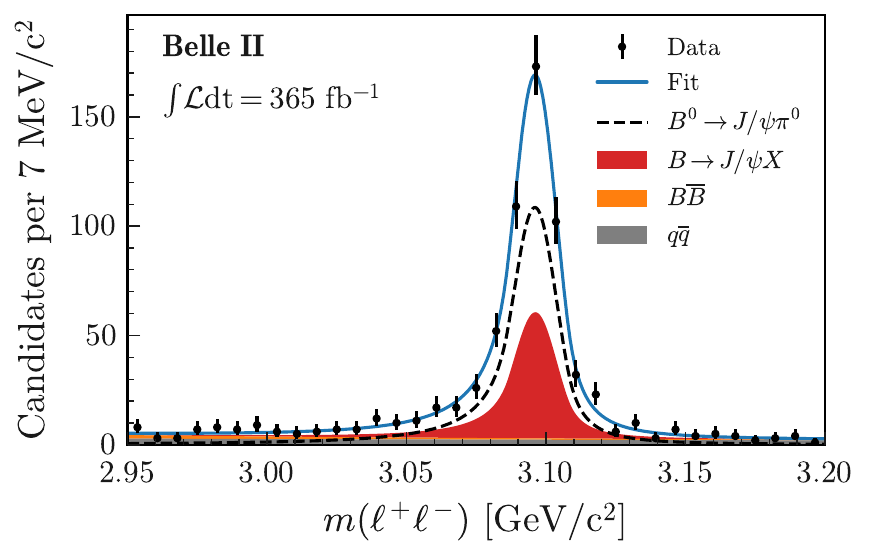}
\caption{Distributions of (top) \deltae and (bottom) \mll for \jpsipiz candidates (data points) with fits overlaid (curves and stacked areas).}
\label{fit:de-fit}
\end{figure}

In the fit to the data, we determine \nsig, \nbkg, and \fbkg, separately for the \jpsiee and \jpsimm modes, while we fix $n_{\BBbar}$ to the values expected in simulation.
The data are displayed in Fig.~\ref{fit:de-fit} with fit projections overlaid.
The signal yields extracted from the fit and signal selection efficiencies are reported in Table~\ref{tab:de-fit}.
We also report the signal purity, defined as $S/(S+B)$, where $S$ and $B$ are the number of signal and background candidates in the signal region.
The latter, which is used to extract the \CP asymmetries, is defined as $|\deltae|<0.1$~\gev and $3.0<\mee<3.14$~\gevcc or $3.025<\mmumu<3.14$~\gevcc.

From the signal yields, we determine the branching fraction
\begin{equation}
\BF = \frac{(\nsig^{\epem}/\eff_{\text{sig}}^{\epem} + \nsig^{\mup\mun}/\eff_{\text{sig}}^{\mup\mun}) (1+\fpm/\fzz)}{\BF(\jpsill)\BF(\piz\to\gamma\gamma)2N(\BBbar)}
\end{equation}
where $\eff_{\text{sig}}$ are the efficiencies obtained from simulated signal samples and corrected for differences between data and simulation using control samples, $\BF(\jpsill)$ is the sum of $\BF(\jpsiee)=(5.971\pm0.032)\%$ and $\BF(\jpsimm)=(5.961\pm0.033)\%$, $\BF(\piz\to\gamma\gamma)$ is $(98.823\pm0.034)\%$~\cite{PhysRevD.110.030001}, $N(\BBbar)=\nbb$ is the number of \BBbar pairs in the dataset, and $\fpm/\fzz=1.052\pm0.031$ is the $\Bp/\Bz$ production ratio~\cite{Banerjee:2024znd}.
We obtain $\BF(\jpsipiz) = (\bfval \pm \bferr)\times 10^{-5}$, where the uncertainty is statistical only.
We validate our analysis on the \jpsikstp control sample, for which we obtain $\BF(\jpsikstp)=\bfkstp$, where the uncertainties are statistical only, in agreement with the world average~\cite{PhysRevD.110.030001}.

\begin{table}[t]
\caption{Signal efficiencies, yields, and purity in the signal region.
The signal efficiencies are corrected for differences between data and simulation, and their uncertainties include both statistical and systematic uncertainties, while the uncertainties on the signal yields are statistical only.}
\label{tab:de-fit}
\begin{tabular*}{\linewidth}{@{\extracolsep{\fill}}lccc}
\hline
\hline
Decay mode  & $\eff_{\text{sig}}$ [\%] & $n_{\text{sig}}$ & Purity [\%] \\
\hline
\jpsimm &$48 \pm 2$ &$204 \pm 17$ &$80$\\
\jpsiee &$41 \pm 2$ &$188 \pm 17$ & $83$\\
\hline
\hline
\end{tabular*}
\end{table}
\section{\CP asymmetry fit}
\label{sec:dt-fit}
We determine the \CP asymmetries from a likelihood fit to the unbinned \deltat distribution of flavor-tagged candidates in the signal region.
Candidates outside of the signal region are removed from the fit as they mostly consist of \jpsix background events which dilute the observable \CP asymmetries of the signal.
The likelihood function is
\begin{equation}
\begin{split}
\lh \propto \prod_{i=1}^{N} \bigg\{
(1-f_{\qqbar}) \Big[
(1-f_{\BBbar})
\big(
(1-f_{\jpsi X})P_{\jpsi \piz}^i \\
+f_{\jpsi X}P_{\jpsi X}^i
\big)
+f_{\BBbar}P_{\BBbar}^i
\Big]
+ f_{\qqbar}P_{\qqbar}^i 
\bigg\}
\end{split}
\label{eq:lh}
\end{equation}
where $f_{\qqbar}$, $f_{\BBbar}$, and $f_{\jpsi X}$ are the background fractions in the signal region, determined from the signal extraction fit, and $P^i=P(\deltat^i)$ is the \pdf of the $i$th candidate.

The \deltat distribution of the signal in Eq.~\ref{eq:dt_theo} is modified to model the effect of imperfect flavor assignment from the tagging algorithm
\begin{equation}
\begin{split}
\mathcal{P}(\dt, q)  = \frac{e^{-|\dt|/\taud}}{4\taud} \Big\{
1-\qtag \dwtag + \qtag \mutag (1-2\wtag) \\
+ \big[\qtag(1-2\wtag)+\mutag(1-q\dwtag)\big] \\
\times \big[  \scp \sin(\dmd\dt) - \ccp \cos(\dmd\dt) \big] \Big\},
\end{split}
\label{eq:dt_tag}
\end{equation}
where $w$ is the wrong-tag probability, $\Delta w$ is the wrong-tag probability difference between events tagged as \Bz and \Bzb, and \mutag is the tagging efficiency asymmetry between \Bz and \Bzb.
We divide our sample into intervals of the tag-quality variable $r=1-2w$ provided by the tagging algorithm, to gain statistical sensitivity from events with different wrong-tag fractions.
We use the boundaries $(0.0,0.1,0.25,0.45,0.6,0.725,0.875,1.0)$ and corresponding calibration parameters obtained with a sample of \dpi decays in Ref.~\cite{PhysRevD.110.012001}.
The effective flavor tagging efficiency, defined as $\sum_{i}\varepsilon_i(1-2w_i)^2$, where $\varepsilon_i$ is the fraction of events associated with a tag decision and $w_i$ is the wrong-tag probability in the $i$-th $r$-bin, is $(37.40\pm0.43\pm0.36)\%$, where the uncertainties are statistical and systematic, respectively~\cite{PhysRevD.110.012001}.
Since $f_{\qqbar}$ varies with $r$, we use the distribution in off-resonance data to scale the average fraction in each bin, while $f_{\BBbar}$ and $f_{\jpsix}$ are constants in $r$, as verified in simulation.

The effect of finite detector \deltat resolution is taken into account by modifying Eq.~\ref{eq:dt_tag} as
\begin{linenomath}
\begin{equation}
\mathcal{F}(\dt,q | \dterr) = \int \mathcal{P}(\dtp, \qtag) \dtres d\dtp,
\label{eq:dt_res}
\end{equation}
\end{linenomath}
where $\mathcal{R}$ is the resolution function, conditional on the per-event \dt uncertainty $\dterr$.
The resolution function is described by the sum of two components:
\begin{linenomath}
\begin{align}
\begin{split}
\mathcal{R}(\delta t | & \sigma_{\dt}) = (1-f_t(\sigma_{\dt}))G(\delta t|m_G\sigma_{\dt}, s_G \sigma_{\dt}) \\
+ &f_t(\sigma_{\dt}) R_t(\delta t| m_t \sigma_{\dt},  s_t \sigma_{\dt}, k/\sigma_{\dt}, f_{>}, f_{<})
\end{split}
\label{eq:reso}
\end{align}
\end{linenomath}
where $\delta t$ is the difference between the observed and the true \dt. 
The first component is a Gaussian function with mean $m_G$ and width $s_G$ scaled by $\sigma_{\Delta t}$, which models the core of the distribution.
The second component $R_t$ is the sum of a Gaussian function and the convolution of a Gaussian with two oppositely sided exponential functions,
\begin{linenomath}
\begin{align}
\begin{split}
R_t(x| \mu, \sigma, k, f_{>}, f_{<}&) = (1-f_{<}-f_{>}) G(x|\mu,\sigma) \\
&+f_{<} G(x|\mu,\sigma) \otimes k \exp_{<}(kx) \\
&+f_{>} G(x|\mu,\sigma) \otimes k \exp_{>}(-kx),
\end{split}
\end{align}
\end{linenomath}
where $\exp_{>}(kx)=\exp(kx)$ if $x>0$ or zero otherwise, and similarly for $\exp_{<}(kx)$.
The exponential tails arise from intermediate displaced charm-hadron vertices from the \btag decay.
The fraction $f_t(\sigma_{\dt})$ is zero at low values of $\sigma_{\dt}$ and rises steeply to reach a plateau of 0.2 at $\sigma_{\dt}=0.25~\ps$.
We neglect an outlier component in the resolution, accounting for $\mathcal{O}(10^{-3})$ fraction of events with poorly reconstructed vertices, which shows no impact on the results.
We use the same resolution function parameters calibrated with a sample of \dpi decays as in Ref.~\cite{PhysRevD.110.012001}.

The \jpsix backgrounds are modeled separately for decays of \Bz and \Bp mesons in the \deltat fit, with effective lifetimes determined from simulation and \pdf with the same functional form as the signal.
The \CP asymmetries of the \bzjpsix backgrounds are determined from simulated \jpsiks and \jpsikl decays misreconstructed as signal.
The fraction of \bpjpsix backgrounds relative to the total amount of \jpsix backgrounds is fixed from simulation.
The \CP asymmetries of the \bpjpsix backgrounds are set to zero.
The \deltat distribution of the \BBbar backgrounds is described using an exponential \pdf convolved with a Gaussian resolution model determined from simulated data. 
The distribution of the continuum background is modeled with a double-Gaussian \pdf determined from off-resonance data. 

We validate the fit by measuring the lifetime on \jpsikstp and \jpsiks data.
We obtain $\tau_{\Bp}=(\taukstp)\ps$ and $\tau_{\Bz}=(\tauks)\ps$, respectively, where the uncertainties are statistical only.
We also perform the \CP asymmetry fit on the \jpsiks sample, for which we obtain $\ccp=\ccpks$ and $\scp=\scpks$, where the uncertainties are statistical only.
All the values of the lifetimes and \CP asymmetries are consistent with the world averages~\cite{PhysRevD.110.030001}.

\begin{figure}
\centering
\includegraphics[width=0.45\textwidth]{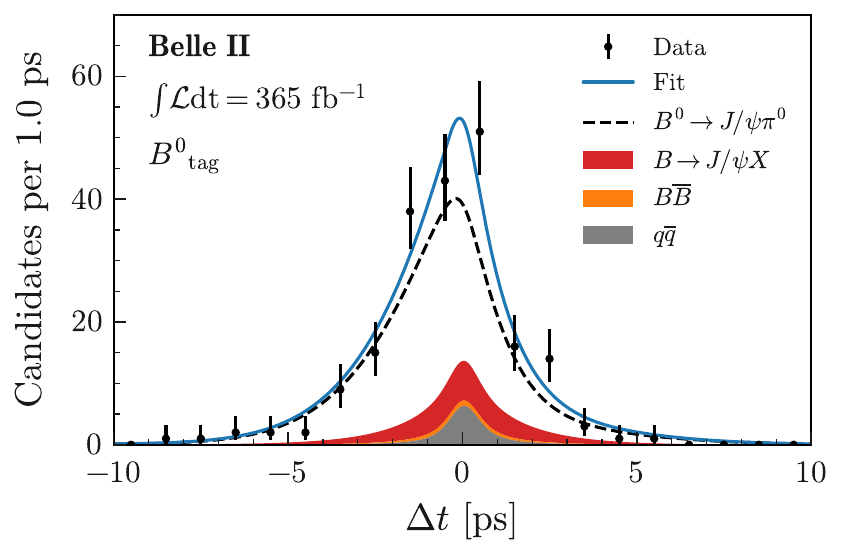}
\includegraphics[width=0.45\textwidth]{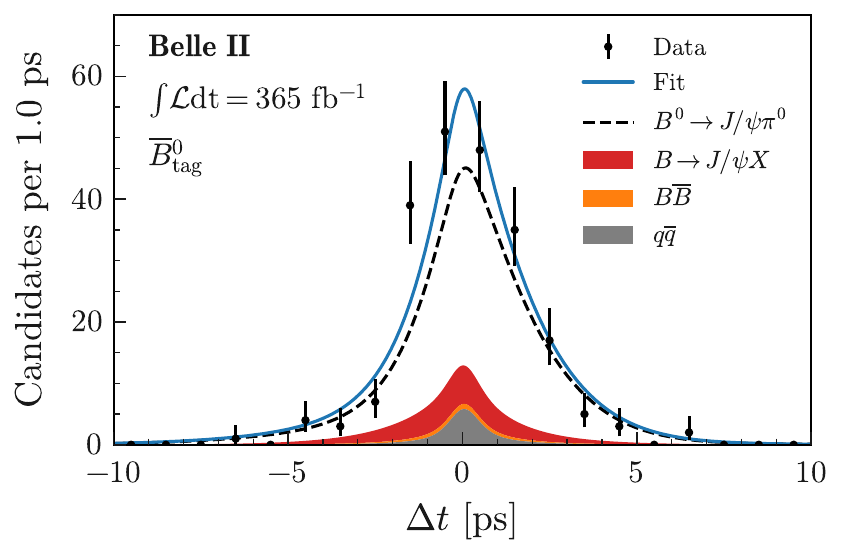}
\caption{Distributions of \deltat for (top) \Bz and (bottom) \Bzb-tagged \jpsipiz candidates (data points) with fits overlaid (curves and stacked areas).}
\label{fit:dt-fit}
\end{figure}

\begin{figure}
\centering
\includegraphics[width=0.48\textwidth]{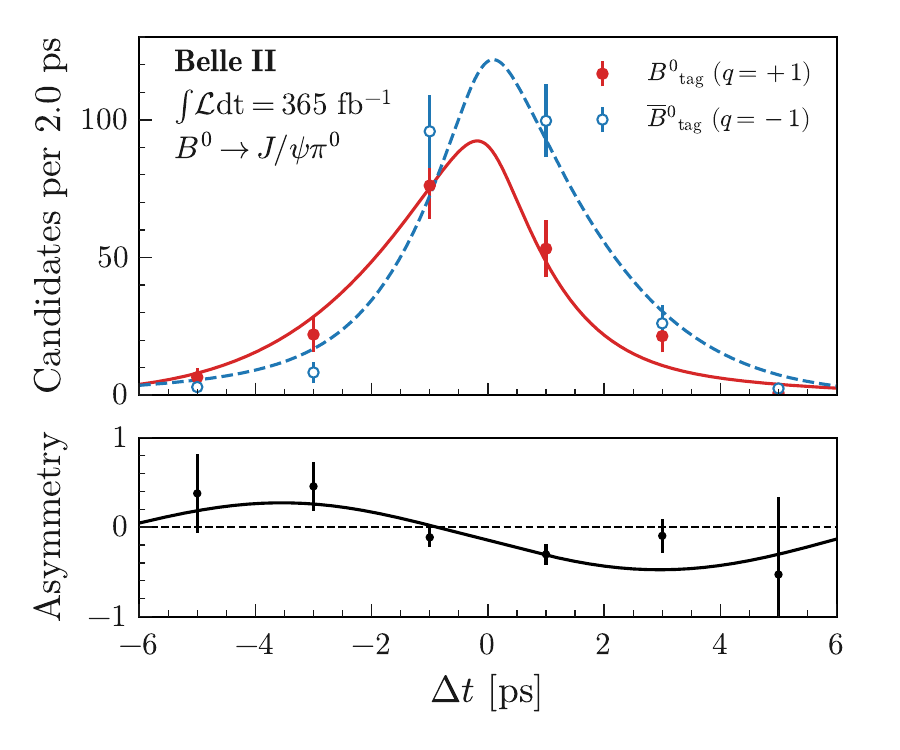}
\caption{Distributions and fit projections of \deltat for background-subtracted flavor-tagged \jpsipiz candidates.
The fit \pdfs corresponding to candidates tagged as $q=-1$ and $q=+$1 are shown as dashed and solid curves, respectively.
The decay rate asymmetry, defined as $[N(q=+1)-N(q=-1)]/[N(q=+1)+N(q=-1)]$, is displayed in the bottom subpanel.}
\label{fit:splot}
\end{figure}

The \jpsipiz data are displayed in Fig.~\ref{fit:dt-fit} with fit projections overlaid.
We determine the direct and mixing-induced \CP asymmetries $\ccp=\ccpval\pm\ccperr$ and $\scp=\scpval\pm\scperr$, where the uncertainties are statistical only.
The correlation between \ccp and \scp is \cpcorr.
The \deltat distributions for tagged signal decays, after subtracting the backgrounds~\cite{Pivk:2004ty}, are displayed in Fig.~\ref{fit:splot}, along with the decay rate asymmetry.

\section{Systematic Uncertainties}
\label{sec:systematics}

Contributions from all considered sources of systematic uncertainty are listed in Tables~\ref{tab:bf-sys} and \ref{tab:cp-sys} for the branching fraction and \CP asymmetries, respectively.
The leading contribution to the total systematic uncertainty on the branching fraction arises from the \piz efficiency calibration, while the main systematic uncertainties on the \CP asymmetries originate from the calibration of the flavor tagging and resolution function with the \dpi control sample and tag-side interference.

\subsection{Branching fraction}
    
In the computation of the branching fraction, we correct the signal efficiencies obtained in simulation using control samples from collision data.
The statistical and systematic uncertainties associated with the correction factors are propagated to the measurement of the branching fraction systematic uncertainty.

The \piz reconstruction efficiency is measured in data and simulation using the ratio of the yields of $\Dstarp\to \Dz(\to \Km \pip \piz)\pip$ and $\Dstarp\to \Dz(\to \Km \pip)\pip$ decays, scaled by the inverse values of their branching fractions.
The yield ratio in experimental and simulated data is used to obtain correction factors as functions of the \piz polar angle and momentum.
The average correction factor over the kinematic distribution of the \piz in \jpsipiz decays is \pizcorr, where the uncertainty is dominated by the knowledge of the \Dz branching fractions~\cite{PhysRevD.110.030001}.

The difference in electron and muon identification performance between simulation and experimental data is calibrated using $\jpsi \to \ell^+\ell^-$, $\epem \to \ell^+\ell^-(\gamma)$ and $\epem\to \epem \ell^+\ell^-$ samples.
The average correction factor over the kinematic distribution of the signal is \eidcorr for the \jpsiee mode and \midcorr for the \jpsimm mode, where the uncertainties are the sum in quadrature of the statistical and systematic uncertainties.

The performance of the continuum-suppression BDT is validated using the \jpsiks control sample.
The ratio of the signal efficiency after applying the BDT requirement in data and simulation is found to be \bdtcorre and \bdtcorrm for the \jpsiee and \jpsimm modes, respectively, where the uncertainties are statistical only.

Tracking efficiencies are measured using $\epem\to \taup\taum$ events, where one $\tau$ decays as $\taum \to e^- \overline{\nu}_e\nu_{\tau}$ and the other as $\taum \to \pim\pip\pim \nu_{\tau}$.
Efficiencies for data and simulation are found to be compatible within an uncertainty of $0.27\%$, which is propagated for each track to the uncertainty on the branching fraction.

\begin{table}[t]
\caption{Relative systematic uncertainties on the branching fraction compared with the statistical uncertainties.}
\label{tab:bf-sys}
\begin{tabular*}{\linewidth}{@{\extracolsep{\fill}}lrr}\hline
\hline
Source & \multicolumn{1}{c}{Relative uncertainty on BF$[\%]$} \\
\hline
\piz efficiency & $ 3.7$\\
Lepton ID & $ 0.4$\\
BDT & $ 0.3$\\
Tracking efficiencies & $ 0.5$\\
External inputs &$ 0.4$\\
$N(\BBbar)$ &$ 1.4$\\
$\fpm/\fzz$ & $ 1.5$\\
Fixed parameters & $ 0.9$\\
Backgrounds composition & $ 0.4$\\
Multiple candidates & $ 0.5$\\
\hline
Total systematic uncertainty & $ 4.5$\\
\hline
Statistical uncertainty & $ 6.0$\\
\hline
\hline
\end{tabular*}
\end{table}

We propagate the uncertainty on the branching fractions of the \jpsi and \piz decay modes used to reconstruct the signal~\cite{PhysRevD.110.030001}.
The uncertainty on the number of \Bz mesons in the sample arises from the measurement of the number of \BBbar pairs and from the knowledge of the $\Bp/\Bz$ production ratio~\cite{Banerjee:2024znd}.
Both uncertainties are propagated to the branching fraction and included in the systematic uncertainty.

We consider the uncertainties associated with the determination of the signal yields from the fit in the following way.
We repeat the fit by fixing the parameters determined in the control samples to alternative values chosen according to their statistical covariance matrix.
We take the standard deviation of the distribution of the signal yields thus obtained and propagate it to the branching fraction. 
To account for differences in the composition of the backgrounds between data and simulation, we use simplified simulated datasets where each component is generated according to their \pdfs.
The main \jpsix background components are generated with independent \deltae distributions and their yields varied between $\pm20\%$ and $\pm50\%$ from the expected value.
We fit these datasets using the nominal fit model and obtain an average bias on the signal yields for each alternative background configuration.
We verify that these variations in the background yields cover possible disagreements between data and simulation by comparing their distributions with sidebands enriched in different type of backgrounds.
We define a sideband with $\mbc^\prime<5.27~\gevcc$ and $\mgg\in[0.1,0.16]~\gevcc$, where the backgrounds with properly reconstructed \piz candidates are dominant, and a sideband  
with $\mbc^\prime>5.27~\gevcc$ and $\mgg\in[0.02,0.1]\cup[0.16,0.2]~\gevcc$, where the backgrounds with mis-reconstructed \piz candidates are dominant.
We also account for variations in the \BBbar background yield and fraction of signal with a misreconstructed \piz using the same approach.
We take the standard deviation of the distribution of the biases as a systematic uncertainty and propagate it to the branching fraction.

Finally, we repeat our measurement on ensembles of simulated data using alternative candidate selection requirements for events with multiple candidates.
For each selection, we obtain an average bias on the signal yields.
We take the standard deviation of the distribution of the average biases as a systematic uncertainty and propagate it to the branching fraction.

\subsection{\CP asymmetries}
We consider the uncertainties associated with the flavor tagging and resolution function calibration.
We repeat the \CP asymmetry fit by fixing the calibration parameters determined in the \dpi sample to alternative values chosen according to their statistical covariance matrix~\cite{PhysRevD.110.012001}.
We also repeat the fit by varying the same set of parameters within their systematic uncertainties without correlations.
In both cases, we take as a systematic uncertainty the standard deviation of the \CP asymmetries distribution thus obtained, and sum them in quadrature.

We propagate the statistical uncertainties on the signal and background fractions to the \CP asymmetries.
We repeat the \CP asymmetry fit by fixing the yields and continuum background fractions determined in the signal extraction fit to alternative values chosen according to their statistical covariance matrix.
The standard deviation of the distribution of the \CP asymmetries thus obtained is assigned as a systematic uncertainty.

\begin{table}[t]
\caption{Systematic uncertainties on the \CP asymmetries compared with the statistical uncertainties.}
\label{tab:cp-sys}
\begin{tabular*}{\linewidth}{@{\extracolsep{\fill}}lrr}\hline
\hline
Source & \ccp & \escp \\
\hline
Calibration with \dpi &  $0.017$ & $0.023$ \\
Signal extraction fit &  $0.003$ & $0.017$ \\
Backgrounds composition &  $0.005$ & $0.009$ \\
Backgrounds \deltat shapes &  $<0.001$ & $0.001$ \\
Fit bias &  $0.010$ & $0.010$ \\
Multiple candidates &  $<0.001$ & $0.002$ \\
Tracking detector misalignment &  $0.002$ & $0.002$ \\
Tag-side interference &  $0.027$ & $0.001$ \\
\taud and \dmd &  $<0.001$ & $<0.001$ \\
\hline
Total systematic uncertainty &  $0.034$ & $0.032$ \\
\hline
Statistical uncertainty &  $0.123$ & $0.171$ \\
\hline
\hline
\end{tabular*}
\end{table}

In order to estimate the systematic uncertainty associated with the background model, we use the ensembles generated with alternative background compositions used for the study of the systematic uncertainties on the branching fraction.
These simplified simulated datasets are also generated with different \deltat distributions for the main \jpsix background components.
In particular, the \jpsiks and \jpsikl backgrounds are generated using the known value of their \CP asymmetries~\cite{PhysRevD.110.030001}.
We generate separately an additional prompt component in \deltat originating from tracks of the signal-side that are included in the fit of the tag-side vertex.
We fit these datasets using the nominal fit model and obtain an average bias on the \CP asymmetries for each alternative background configuration.
We take the standard deviation of the distribution of these biases as a systematic uncertainty.

We also consider the variations of the parameters of the \deltat \pdf of the continuum background using the covariance matrix determined in the fit to the off-resonance data.
We take as systematic uncertainty the standard deviation of the distribution of the \CP asymmetries thus obtained.

We estimate a fit bias, due to the combined effects of the approximate determination of \deltat in Eq.~\ref{eq:dt-corr} and differences between the signal and calibration sample, using simulated signal events generated with  \ccp in $[-0.4,0.4]$ and \scp in $[-1.0,0.0]$ in steps of $0.2$.
In the nominal fit to the data, we correct the \CP asymmetries for their bias (\ccpbias on \ccp and \scpbias on \scp, where the uncertainties come from the size of the simulated sample) and assign the absolute value of the bias (0.01) as a systematic uncertainty.

The same procedure used to estimate the impact of the candidate selection on the measurement of the branching fraction is repeated for the \CP asymmetries.

We study the impact of the tracking detector misalignment on the \CP asymmetries using simulated samples reconstructed with various misalignment configurations and assign as a systematic uncertainty the sum in quadrature of the differences with respect to the nominal alignment configuration.

We estimate the shift from the true values of the \CP asymmetries due to the tag-side interference, \ie,~neglecting the effect of CKM-suppressed $\bquark \to \uquark\cquarkbar\dquark$ decays in the \btag in the model for \dt, using the estimators for \ccp and \scp that reproduce this bias as given in Ref.~\cite{PhysRevD.68.034010}.
Since the sign of the bias depends on the strong phase difference between the favored and suppressed decays, which is poorly known, we take the maximum absolute value as a systematic uncertainty.

The values of the \Bz lifetime and oscillation frequency are fixed in the \pdf. 
To estimate the corresponding systematic uncertainties, we vary them around their known values according to their uncertainties~\cite{PhysRevD.110.030001}.
We find that this has negligible impact on the \CP asymmetries.
\section{Summary}
\label{sec:summary}

We report a measurement of the branching fraction and \CP asymmetries in \jpsipiz decays using data from the \belletwo experiment.
We find \njpsipiz signal decays in a sample containing \nbb \BBbar events, corresponding to a value of the branching fraction of
\begin{equation}
\bf(\jpsipiz) = (\bfval \pm \bferr \pm \bfsys)\times 10^{-5},
\end{equation}
where the first uncertainty is statistical, and the second is systematic.
The result is consistent with the world average~\cite{PhysRevD.110.030001} and has comparable precision to previous determinations.

We obtain the following values of the \CP asymmetries
\begin{equation}
\begin{split}
\ccp &= \ccpval \pm \ccperr \pm \ccpsys, \\
\scp &= \scpval \pm \scperr \pm \scpsys,
\end{split}
\end{equation}
where the first uncertainty is statistical, and the second is systematic.
The results are the most precise to date and are consistent with previous determinations from \belle and \babar~\cite{BaBar:2008kfx,Belle:2018nxw}.
The central value of \scp is $5.0$ standard deviations from zero.
The significance is calculated using the sum in quadrature of the statistical and systematic uncertainties.
This is the first observation of mixing-induced \CP violation in \jpsipiz decays from a single measurement.
The improved determinations of the branching fraction and \CP asymmetries in this mode provide further constraints on the penguin parameters and on the extraction of the CKM angle $\phi_1$~\cite{Barel_2021}.

\section*{Acknowledgements}

This work, based on data collected using the Belle II detector, which was built and commissioned prior to March 2019,
was supported by
Higher Education and Science Committee of the Republic of Armenia Grant No.~23LCG-1C011;
Australian Research Council and Research Grants
No.~DP200101792, 
No.~DP210101900, 
No.~DP210102831, 
No.~DE220100462, 
No.~LE210100098, 
and
No.~LE230100085; 
Austrian Federal Ministry of Education, Science and Research,
Austrian Science Fund
No.~P~34529,
No.~J~4731,
No.~J~4625,
and
No.~M~3153,
and
Horizon 2020 ERC Starting Grant No.~947006 ``InterLeptons'';
Natural Sciences and Engineering Research Council of Canada, Compute Canada and CANARIE;
National Key R\&D Program of China under Contract No.~2022YFA1601903,
National Natural Science Foundation of China and Research Grants
No.~11575017,
No.~11761141009,
No.~11705209,
No.~11975076,
No.~12135005,
No.~12150004,
No.~12161141008,
and
No.~12175041,
and Shandong Provincial Natural Science Foundation Project~ZR2022JQ02;
the Czech Science Foundation Grant No.~22-18469S 
and
Charles University Grant Agency project No.~246122;
European Research Council, Seventh Framework PIEF-GA-2013-622527,
Horizon 2020 ERC-Advanced Grants No.~267104 and No.~884719,
Horizon 2020 ERC-Consolidator Grant No.~819127,
Horizon 2020 Marie Sklodowska-Curie Grant Agreement No.~700525 ``NIOBE''
and
No.~101026516,
and
Horizon 2020 Marie Sklodowska-Curie RISE project JENNIFER2 Grant Agreement No.~822070 (European grants);
L'Institut National de Physique Nucl\'{e}aire et de Physique des Particules (IN2P3) du CNRS
and
L'Agence Nationale de la Recherche (ANR) under grant ANR-21-CE31-0009 (France);
BMBF, DFG, HGF, MPG, and AvH Foundation (Germany);
Department of Atomic Energy under Project Identification No.~RTI 4002,
Department of Science and Technology,
and
UPES SEED funding programs
No.~UPES/R\&D-SEED-INFRA/17052023/01 and
No.~UPES/R\&D-SOE/20062022/06 (India);
Israel Science Foundation Grant No.~2476/17,
U.S.-Israel Binational Science Foundation Grant No.~2016113, and
Israel Ministry of Science Grant No.~3-16543;
Istituto Nazionale di Fisica Nucleare and the Research Grants BELLE2;
Japan Society for the Promotion of Science, Grant-in-Aid for Scientific Research Grants
No.~16H03968,
No.~16H03993,
No.~16H06492,
No.~16K05323,
No.~17H01133,
No.~17H05405,
No.~18K03621,
No.~18H03710,
No.~18H05226,
No.~19H00682, 
No.~20H05850,
No.~20H05858,
No.~22H00144,
No.~22K14056,
No.~22K21347,
No.~23H05433,
No.~26220706,
and
No.~26400255,
and
the Ministry of Education, Culture, Sports, Science, and Technology (MEXT) of Japan;  
National Research Foundation (NRF) of Korea Grants
No.~2016R1-D1A1B-02012900,
No.~2018R1-A6A1A-06024970,
No.~2021R1-A6A1A-03043957,
No.~2021R1-F1A-1060423,
No.~2021R1-F1A-1064008,
No.~2022R1-A2C-1003993,
No.~2022R1-A2C-1092335,
No.~RS-2023-00208693,
No.~RS-2024-00354342
and
No.~RS-2022-00197659,
Radiation Science Research Institute,
Foreign Large-Size Research Facility Application Supporting project,
the Global Science Experimental Data Hub Center, the Korea Institute of
Science and Technology Information (K24L2M1C4)
and
KREONET/GLORIAD;
Universiti Malaya RU grant, Akademi Sains Malaysia, and Ministry of Education Malaysia;
Frontiers of Science Program Contracts
No.~FOINS-296,
No.~CB-221329,
No.~CB-236394,
No.~CB-254409,
and
No.~CB-180023, and SEP-CINVESTAV Research Grant No.~237 (Mexico);
the Polish Ministry of Science and Higher Education and the National Science Center;
the Ministry of Science and Higher Education of the Russian Federation
and
the HSE University Basic Research Program, Moscow;
University of Tabuk Research Grants
No.~S-0256-1438 and No.~S-0280-1439 (Saudi Arabia);
Slovenian Research Agency and Research Grants
No.~J1-9124
and
No.~P1-0135;
Agencia Estatal de Investigacion, Spain
Grant No.~RYC2020-029875-I
and
Generalitat Valenciana, Spain
Grant No.~CIDEGENT/2018/020;
The Knut and Alice Wallenberg Foundation (Sweden), Contracts No.~2021.0174 and No.~2021.0299;
National Science and Technology Council,
and
Ministry of Education (Taiwan);
Thailand Center of Excellence in Physics;
TUBITAK ULAKBIM (Turkey);
National Research Foundation of Ukraine, Project No.~2020.02/0257,
and
Ministry of Education and Science of Ukraine;
the U.S. National Science Foundation and Research Grants
No.~PHY-1913789 
and
No.~PHY-2111604, 
and the U.S. Department of Energy and Research Awards
No.~DE-AC06-76RLO1830, 
No.~DE-SC0007983, 
No.~DE-SC0009824, 
No.~DE-SC0009973, 
No.~DE-SC0010007, 
No.~DE-SC0010073, 
No.~DE-SC0010118, 
No.~DE-SC0010504, 
No.~DE-SC0011784, 
No.~DE-SC0012704, 
No.~DE-SC0019230, 
No.~DE-SC0021274, 
No.~DE-SC0021616, 
No.~DE-SC0022350, 
No.~DE-SC0023470; 
and
the Vietnam Academy of Science and Technology (VAST) under Grants
No.~NVCC.05.12/22-23
and
No.~DL0000.02/24-25.

These acknowledgements are not to be interpreted as an endorsement of any statement made
by any of our institutes, funding agencies, governments, or their representatives.

We thank the SuperKEKB team for delivering high-luminosity collisions;
the KEK cryogenics group for the efficient operation of the detector solenoid magnet and IBBelle on site;
the KEK Computer Research Center for on-site computing support; the NII for SINET6 network support;
and the raw-data centers hosted by BNL, DESY, GridKa, IN2P3, INFN, 
and the University of Victoria.

\bibliographystyle{apsrev4-1}
\bibliography{references}

\begin{thebibliography}{51}%
\makeatletter
\providecommand \@ifxundefined [1]{%
 \@ifx{#1\undefined}
}%
\providecommand \@ifnum [1]{%
 \ifnum #1\expandafter \@firstoftwo
 \else \expandafter \@secondoftwo
 \fi
}%
\providecommand \@ifx [1]{%
 \ifx #1\expandafter \@firstoftwo
 \else \expandafter \@secondoftwo
 \fi
}%
\providecommand \natexlab [1]{#1}%
\providecommand \enquote  [1]{``#1''}%
\providecommand \bibnamefont  [1]{#1}%
\providecommand \bibfnamefont [1]{#1}%
\providecommand \citenamefont [1]{#1}%
\providecommand \href@noop [0]{\@secondoftwo}%
\providecommand \href [0]{\begingroup \@sanitize@url \@href}%
\providecommand \@href[1]{\@@startlink{#1}\@@href}%
\providecommand \@@href[1]{\endgroup#1\@@endlink}%
\providecommand \@sanitize@url [0]{\catcode `\\12\catcode `\$12\catcode
  `\&12\catcode `\#12\catcode `\^12\catcode `\_12\catcode `\%12\relax}%
\providecommand \@@startlink[1]{}%
\providecommand \@@endlink[0]{}%
\providecommand \url  [0]{\begingroup\@sanitize@url \@url }%
\providecommand \@url [1]{\endgroup\@href {#1}{\urlprefix }}%
\providecommand \urlprefix  [0]{URL }%
\providecommand \Eprint [0]{\href }%
\providecommand \doibase [0]{http://dx.doi.org/}%
\providecommand \selectlanguage [0]{\@gobble}%
\providecommand \bibinfo  [0]{\@secondoftwo}%
\providecommand \bibfield  [0]{\@secondoftwo}%
\providecommand \translation [1]{[#1]}%
\providecommand \BibitemOpen [0]{}%
\providecommand \bibitemStop [0]{}%
\providecommand \bibitemNoStop [0]{.\EOS\space}%
\providecommand \EOS [0]{\spacefactor3000\relax}%
\providecommand \BibitemShut  [1]{\csname bibitem#1\endcsname}%
\let\auto@bib@innerbib\@empty
\bibitem [{\citenamefont {Cabibbo}(1963)}]{PhysRevLett.10.531}%
  \BibitemOpen
  \bibfield  {author} {\bibinfo {author} {\bibfnamefont {N.}~\bibnamefont
  {Cabibbo}},\ }\href {\doibase 10.1103/PhysRevLett.10.531} {\bibfield
  {journal} {\bibinfo  {journal} {Phys. Rev. Lett.}\ }\textbf {\bibinfo
  {volume} {10}},\ \bibinfo {pages} {531} (\bibinfo {year} {1963})}\BibitemShut
  {NoStop}%
\bibitem [{\citenamefont {Kobayashi}\ and\ \citenamefont
  {Maskawa}(1973)}]{Kobayashi:1973fv}%
  \BibitemOpen
  \bibfield  {author} {\bibinfo {author} {\bibfnamefont {M.}~\bibnamefont
  {Kobayashi}}\ and\ \bibinfo {author} {\bibfnamefont {T.}~\bibnamefont
  {Maskawa}},\ }\href {\doibase 10.1143/PTP.49.652} {\bibfield  {journal}
  {\bibinfo  {journal} {Prog. Theor. Phys.}\ }\textbf {\bibinfo {volume}
  {49}},\ \bibinfo {pages} {652} (\bibinfo {year} {1973})}\BibitemShut
  {NoStop}%
\bibitem [{con()}]{convention}%
  \BibitemOpen
  \href@noop {} {}\bibinfo {howpublished} {{This angle is also known as
  $\beta$.}}\BibitemShut {Stop}%
\bibitem [{\citenamefont {Aubert}\ \emph {et~al.}(2009)\citenamefont {Aubert}
  \emph {et~al.}}]{PhysRevD.79.072009}%
  \BibitemOpen
  \bibfield  {author} {\bibinfo {author} {\bibfnamefont {B.}~\bibnamefont
  {Aubert}} \emph {et~al.} (\bibinfo {collaboration} {\babar Collaboration}),\
  }\href {\doibase 10.1103/PhysRevD.79.072009} {\bibfield  {journal} {\bibinfo
  {journal} {Phys. Rev. D}\ }\textbf {\bibinfo {volume} {79}},\ \bibinfo
  {pages} {072009} (\bibinfo {year} {2009})}\BibitemShut {NoStop}%
\bibitem [{\citenamefont {Adachi}\ \emph {et~al.}(2012)\citenamefont {Adachi}
  \emph {et~al.}}]{PhysRevLett.108.171802}%
  \BibitemOpen
  \bibfield  {author} {\bibinfo {author} {\bibfnamefont {I.}~\bibnamefont
  {Adachi}} \emph {et~al.} (\bibinfo {collaboration} {Belle Collaboration}),\
  }\href {\doibase 10.1103/PhysRevLett.108.171802} {\bibfield  {journal}
  {\bibinfo  {journal} {Phys. Rev. Lett.}\ }\textbf {\bibinfo {volume} {108}},\
  \bibinfo {pages} {171802} (\bibinfo {year} {2012})}\BibitemShut {NoStop}%
\bibitem [{\citenamefont {Aaij}\ \emph
  {et~al.}(2024{\natexlab{a}})\citenamefont {Aaij} \emph
  {et~al.}}]{PhysRevLett.132.021801}%
  \BibitemOpen
  \bibfield  {author} {\bibinfo {author} {\bibfnamefont {R.}~\bibnamefont
  {Aaij}} \emph {et~al.} (\bibinfo {collaboration} {LHCb Collaboration}),\
  }\href {\doibase 10.1103/PhysRevLett.132.021801} {\bibfield  {journal}
  {\bibinfo  {journal} {Phys. Rev. Lett.}\ }\textbf {\bibinfo {volume} {132}},\
  \bibinfo {pages} {021801} (\bibinfo {year} {2024}{\natexlab{a}})}\BibitemShut
  {NoStop}%
\bibitem [{\citenamefont {Grossman}\ and\ \citenamefont
  {Worah}(1997)}]{GROSSMAN1997241}%
  \BibitemOpen
  \bibfield  {author} {\bibinfo {author} {\bibfnamefont {Y.}~\bibnamefont
  {Grossman}}\ and\ \bibinfo {author} {\bibfnamefont {M.~P.}\ \bibnamefont
  {Worah}},\ }\href {\doibase https://doi.org/10.1016/S0370-2693(97)00068-3}
  {\bibfield  {journal} {\bibinfo  {journal} {Physics Letters B}\ }\textbf
  {\bibinfo {volume} {395}},\ \bibinfo {pages} {241} (\bibinfo {year}
  {1997})}\BibitemShut {NoStop}%
\bibitem [{\citenamefont {Amhis}\ \emph {et~al.}(2023)\citenamefont {Amhis}
  \emph {et~al.}}]{HeavyFlavorAveragingGroup:2022wzx}%
  \BibitemOpen
  \bibfield  {author} {\bibinfo {author} {\bibfnamefont {Y.~S.}\ \bibnamefont
  {Amhis}} \emph {et~al.} (\bibinfo {collaboration} {{HFLAV Collaboration}}),\
  }\href {\doibase https://doi.org/10.1103/PhysRevD.107.052008} {\bibfield
  {journal} {\bibinfo  {journal} {Phys. Rev. D}\ }\textbf {\bibinfo {volume}
  {107}},\ \bibinfo {pages} {052008} (\bibinfo {year} {2023})}\BibitemShut
  {NoStop}%
\bibitem [{\citenamefont {Charles}\ \emph {et~al.}(2005)\citenamefont {Charles}
  \emph {et~al.}}]{Charles:2004jd}%
  \BibitemOpen
  \bibfield  {author} {\bibinfo {author} {\bibfnamefont {J.}~\bibnamefont
  {Charles}} \emph {et~al.} (\bibinfo {collaboration} {CKMfitter Group}),\
  }\href {\doibase 10.1140/epjc/s2005-02169-1} {\bibfield  {journal} {\bibinfo
  {journal} {Eur. Phys. J. C}\ }\textbf {\bibinfo {volume} {41}},\ \bibinfo
  {pages} {1} (\bibinfo {year} {2005})},\ \bibinfo {note} {updated results and
  plots available at: \url{http://ckmfitter.in2p3.fr}}\BibitemShut {NoStop}%
\bibitem [{\citenamefont {Bona}\ \emph {et~al.}(2005)\citenamefont {Bona} \emph
  {et~al.}}]{UTfit:2005ras}%
  \BibitemOpen
  \bibfield  {author} {\bibinfo {author} {\bibfnamefont {M.}~\bibnamefont
  {Bona}} \emph {et~al.} (\bibinfo {collaboration} {UTfit Collaboration}),\
  }\href {\doibase 10.1088/1126-6708/2005/07/028} {\bibfield  {journal}
  {\bibinfo  {journal} {JHEP}\ }\textbf {\bibinfo {volume} {07}},\ \bibinfo
  {pages} {028} (\bibinfo {year} {2005})},\ \bibinfo {note} {see also online
  updates at \url{http://www.utfit.org}}\BibitemShut {NoStop}%
\bibitem [{\citenamefont {Ciuchini}\ \emph {et~al.}(2005)\citenamefont
  {Ciuchini}, \citenamefont {Pierini},\ and\ \citenamefont
  {Silvestrini}}]{Ciuchini:2005mg}%
  \BibitemOpen
  \bibfield  {author} {\bibinfo {author} {\bibfnamefont {M.}~\bibnamefont
  {Ciuchini}}, \bibinfo {author} {\bibfnamefont {M.}~\bibnamefont {Pierini}}, \
  and\ \bibinfo {author} {\bibfnamefont {L.}~\bibnamefont {Silvestrini}},\
  }\href {\doibase 10.1103/PhysRevLett.95.221804} {\bibfield  {journal}
  {\bibinfo  {journal} {Phys. Rev. Lett.}\ }\textbf {\bibinfo {volume} {95}},\
  \bibinfo {pages} {221804} (\bibinfo {year} {2005})}\BibitemShut {NoStop}%
\bibitem [{\citenamefont {Barel}\ \emph {et~al.}(2021)\citenamefont {Barel},
  \citenamefont {Bruyn}, \citenamefont {Fleischer},\ and\ \citenamefont
  {Malami}}]{Barel_2021}%
  \BibitemOpen
  \bibfield  {author} {\bibinfo {author} {\bibfnamefont {M.~Z.}\ \bibnamefont
  {Barel}}, \bibinfo {author} {\bibfnamefont {K.~D.}\ \bibnamefont {Bruyn}},
  \bibinfo {author} {\bibfnamefont {R.}~\bibnamefont {Fleischer}}, \ and\
  \bibinfo {author} {\bibfnamefont {E.}~\bibnamefont {Malami}},\ }\href
  {\doibase 10.1088/1361-6471/abf2a2} {\bibfield  {journal} {\bibinfo
  {journal} {Journal of Physics G: Nuclear and Particle Physics}\ }\textbf
  {\bibinfo {volume} {48}},\ \bibinfo {pages} {065002} (\bibinfo {year}
  {2021})}\BibitemShut {NoStop}%
\bibitem [{\citenamefont {Barel}\ \emph {et~al.}(2023)\citenamefont {Barel},
  \citenamefont {De~Bruyn}, \citenamefont {Fleischer},\ and\ \citenamefont
  {Malami}}]{Barel:2023oa}%
  \BibitemOpen
  \bibfield  {author} {\bibinfo {author} {\bibfnamefont {M.}~\bibnamefont
  {Barel}}, \bibinfo {author} {\bibfnamefont {K.}~\bibnamefont {De~Bruyn}},
  \bibinfo {author} {\bibfnamefont {R.}~\bibnamefont {Fleischer}}, \ and\
  \bibinfo {author} {\bibfnamefont {E.}~\bibnamefont {Malami}},\ }\href
  {\doibase 10.22323/1.411.0111} {\bibfield  {journal} {\bibinfo  {journal}
  {PoS}\ }\textbf {\bibinfo {volume} {CKM2021}},\ \bibinfo {pages} {111}
  (\bibinfo {year} {2023})}\BibitemShut {NoStop}%
\bibitem [{\citenamefont {Aubert}\ \emph {et~al.}(2008)\citenamefont {Aubert}
  \emph {et~al.}}]{BaBar:2008kfx}%
  \BibitemOpen
  \bibfield  {author} {\bibinfo {author} {\bibfnamefont {B.}~\bibnamefont
  {Aubert}} \emph {et~al.} (\bibinfo {collaboration} {\babar Collaboration}),\
  }\href {\doibase 10.1103/PhysRevLett.101.021801} {\bibfield  {journal}
  {\bibinfo  {journal} {Phys. Rev. Lett.}\ }\textbf {\bibinfo {volume} {101}},\
  \bibinfo {pages} {021801} (\bibinfo {year} {2008})}\BibitemShut {NoStop}%
\bibitem [{\citenamefont {Pal}\ \emph {et~al.}(2018)\citenamefont {Pal} \emph
  {et~al.}}]{Belle:2018nxw}%
  \BibitemOpen
  \bibfield  {author} {\bibinfo {author} {\bibfnamefont {B.}~\bibnamefont
  {Pal}} \emph {et~al.} (\bibinfo {collaboration} {Belle Collaboration}),\
  }\href@noop {} {\bibfield  {journal} {\bibinfo  {journal} {Phys. Rev. D}\
  }\textbf {\bibinfo {volume} {98}},\ \bibinfo {pages} {112008} (\bibinfo
  {year} {2018})}\BibitemShut {NoStop}%
\bibitem [{\citenamefont {Avery}\ \emph {et~al.}(2000)\citenamefont {Avery}
  \emph {et~al.}}]{PhysRevD.62.051101}%
  \BibitemOpen
  \bibfield  {author} {\bibinfo {author} {\bibfnamefont {P.}~\bibnamefont
  {Avery}} \emph {et~al.} (\bibinfo {collaboration} {CLEO Collaboration}),\
  }\href {\doibase 10.1103/PhysRevD.62.051101} {\bibfield  {journal} {\bibinfo
  {journal} {Phys. Rev. D}\ }\textbf {\bibinfo {volume} {62}},\ \bibinfo
  {pages} {051101} (\bibinfo {year} {2000})}\BibitemShut {NoStop}%
\bibitem [{\citenamefont {Navas}\ \emph {et~al.}(2024)\citenamefont {Navas}
  \emph {et~al.}}]{PhysRevD.110.030001}%
  \BibitemOpen
  \bibfield  {author} {\bibinfo {author} {\bibfnamefont {S.}~\bibnamefont
  {Navas}} \emph {et~al.} (\bibinfo {collaboration} {Particle Data Group
  Collaboration}),\ }\href {\doibase 10.1103/PhysRevD.110.030001} {\bibfield
  {journal} {\bibinfo  {journal} {Phys. Rev. D}\ }\textbf {\bibinfo {volume}
  {110}},\ \bibinfo {pages} {030001} (\bibinfo {year} {2024})}\BibitemShut
  {NoStop}%
\bibitem [{\citenamefont {Aaij}\ \emph
  {et~al.}(2024{\natexlab{b}})\citenamefont {Aaij} \emph
  {et~al.}}]{LHCb:2024ier}%
  \BibitemOpen
  \bibfield  {author} {\bibinfo {author} {\bibfnamefont {R.}~\bibnamefont
  {Aaij}} \emph {et~al.} (\bibinfo {collaboration} {LHCb Collaboration}),\
  }\href {\doibase 10.1007/JHEP05(2024)065} {\bibfield  {journal} {\bibinfo
  {journal} {JHEP}\ }\textbf {\bibinfo {volume} {05}},\ \bibinfo {pages} {065}
  (\bibinfo {year} {2024}{\natexlab{b}})}\BibitemShut {NoStop}%
\bibitem [{\citenamefont {Akai}\ \emph {et~al.}(2018)\citenamefont {Akai},
  \citenamefont {Furukawa},\ and\ \citenamefont {Koiso}}]{Akai:2018mbz}%
  \BibitemOpen
  \bibfield  {author} {\bibinfo {author} {\bibfnamefont {K.}~\bibnamefont
  {Akai}}, \bibinfo {author} {\bibfnamefont {K.}~\bibnamefont {Furukawa}}, \
  and\ \bibinfo {author} {\bibfnamefont {H.}~\bibnamefont {Koiso}},\ }\href
  {\doibase 10.1016/j.nima.2018.08.017} {\bibfield  {journal} {\bibinfo
  {journal} {Nucl. Instrum. Meth.}\ }\textbf {\bibinfo {volume} {A907}},\
  \bibinfo {pages} {188} (\bibinfo {year} {2018})}\BibitemShut {NoStop}%
\bibitem [{\citenamefont {Abe}\ \emph {et~al.}()\citenamefont {Abe} \emph
  {et~al.}}]{Abe:2010gxa}%
  \BibitemOpen
  \bibfield  {author} {\bibinfo {author} {\bibfnamefont {T.}~\bibnamefont
  {Abe}} \emph {et~al.} (\bibinfo {collaboration} {Belle II Collaboration}),\
  }\href@noop {} {\ }\Eprint {http://arxiv.org/abs/1011.0352} {arXiv:1011.0352}
  \BibitemShut {NoStop}%
\bibitem [{\citenamefont {Adachi}\ \emph {et~al.}(2025)\citenamefont {Adachi}
  \emph {et~al.}}]{Belle-II:2024vuc}%
  \BibitemOpen
  \bibfield  {author} {\bibinfo {author} {\bibfnamefont {I.}~\bibnamefont
  {Adachi}} \emph {et~al.} (\bibinfo {collaboration} {Belle~II
  Collaboration}),\ }\href@noop {} {\bibfield  {journal} {\bibinfo  {journal}
  {Chin. Phys. C}\ }\textbf {\bibinfo {volume} {49}},\ \bibinfo {pages}
  {013001} (\bibinfo {year} {2025})}\BibitemShut {NoStop}%
\bibitem [{\citenamefont {Adachi}\ \emph {et~al.}(2024)\citenamefont {Adachi}
  \emph {et~al.}}]{PhysRevD.110.012001}%
  \BibitemOpen
  \bibfield  {author} {\bibinfo {author} {\bibfnamefont {I.}~\bibnamefont
  {Adachi}} \emph {et~al.} (\bibinfo {collaboration} {Belle II
  Collaboration}),\ }\href {\doibase 10.1103/PhysRevD.110.012001} {\bibfield
  {journal} {\bibinfo  {journal} {Phys. Rev. D}\ }\textbf {\bibinfo {volume}
  {110}},\ \bibinfo {pages} {012001} (\bibinfo {year} {2024})}\BibitemShut
  {NoStop}%
\bibitem [{\citenamefont {Adamczyk}\ \emph {et~al.}(2022)\citenamefont
  {Adamczyk} \emph {et~al.}}]{Belle-IISVD:2022upf}%
  \BibitemOpen
  \bibfield  {author} {\bibinfo {author} {\bibfnamefont {K.}~\bibnamefont
  {Adamczyk}} \emph {et~al.} (\bibinfo {collaboration} {Belle II SVD
  Collaboration}),\ }\href {\doibase 10.1088/1748-0221/17/11/P11042} {\bibfield
   {journal} {\bibinfo  {journal} {J. Instrum.}\ }\textbf {\bibinfo {volume}
  {17}},\ \bibinfo {pages} {P11042} (\bibinfo {year} {2022})}\BibitemShut
  {NoStop}%
\bibitem [{\citenamefont {Jadach}\ \emph {et~al.}(2000)\citenamefont {Jadach},
  \citenamefont {Ward},\ and\ \citenamefont {W\c{a}s}}]{Jadach:1999vf}%
  \BibitemOpen
  \bibfield  {author} {\bibinfo {author} {\bibfnamefont {S.}~\bibnamefont
  {Jadach}}, \bibinfo {author} {\bibfnamefont {B.~F.~L.}\ \bibnamefont {Ward}},
  \ and\ \bibinfo {author} {\bibfnamefont {Z.}~\bibnamefont {W\c{a}s}},\ }\href
  {\doibase 10.1016/S0010-4655(00)00048-5} {\bibfield  {journal} {\bibinfo
  {journal} {Comput. Phys. Commun.}\ }\textbf {\bibinfo {volume} {130}},\
  \bibinfo {pages} {260} (\bibinfo {year} {2000})}\BibitemShut {NoStop}%
\bibitem [{\citenamefont {Sj\"{o}strand}\ \emph {et~al.}(2015)\citenamefont
  {Sj\"{o}strand} \emph {et~al.}}]{Sjostrand:2014zea}%
  \BibitemOpen
  \bibfield  {author} {\bibinfo {author} {\bibfnamefont {T.}~\bibnamefont
  {Sj\"{o}strand}} \emph {et~al.},\ }\href {\doibase 10.1016/j.cpc.2015.01.024}
  {\bibfield  {journal} {\bibinfo  {journal} {Comput. Phys. Commun.}\ }\textbf
  {\bibinfo {volume} {191}},\ \bibinfo {pages} {159} (\bibinfo {year}
  {2015})}\BibitemShut {NoStop}%
\bibitem [{\citenamefont {Lange}(2001)}]{Lange:2001uf}%
  \BibitemOpen
  \bibfield  {author} {\bibinfo {author} {\bibfnamefont {D.~J.}\ \bibnamefont
  {Lange}},\ }\href {\doibase 10.1016/S0168-9002(01)00089-4} {\bibfield
  {journal} {\bibinfo  {journal} {Nucl. Instrum. Methods Phys. Res., Sect A}\
  }\textbf {\bibinfo {volume} {{462}}},\ \bibinfo {pages} {152} (\bibinfo
  {year} {2001})}\BibitemShut {NoStop}%
\bibitem [{\citenamefont {Agostinelli}\ \emph {et~al.}(2003)\citenamefont
  {Agostinelli} \emph {et~al.}}]{Agostinelli:2002hh}%
  \BibitemOpen
  \bibfield  {author} {\bibinfo {author} {\bibfnamefont {S.}~\bibnamefont
  {Agostinelli}} \emph {et~al.} (\bibinfo {collaboration} {GEANT4
  Collaboration}),\ }\href {\doibase 10.1016/S0168-9002(03)01368-8} {\bibfield
  {journal} {\bibinfo  {journal} {Nucl. Instrum. Methods Phys. Res., Sect. A}\
  }\textbf {\bibinfo {volume} {{506}}},\ \bibinfo {pages} {250} (\bibinfo
  {year} {2003})}\BibitemShut {NoStop}%
\bibitem [{\citenamefont {Lewis}\ \emph {et~al.}(2019)\citenamefont {Lewis}
  \emph {et~al.}}]{Lewis:2018ayu}%
  \BibitemOpen
  \bibfield  {author} {\bibinfo {author} {\bibfnamefont {P.~M.}\ \bibnamefont
  {Lewis}} \emph {et~al.},\ }\href {\doibase 10.1016/j.nima.2018.05.071}
  {\bibfield  {journal} {\bibinfo  {journal} {Nucl. Instrum. Meth. A}\ }\textbf
  {\bibinfo {volume} {914}},\ \bibinfo {pages} {69} (\bibinfo {year}
  {2019})}\BibitemShut {NoStop}%
\bibitem [{\citenamefont {Liptak}\ \emph {et~al.}(2022)\citenamefont {Liptak}
  \emph {et~al.}}]{Liptak:2021tog}%
  \BibitemOpen
  \bibfield  {author} {\bibinfo {author} {\bibfnamefont {Z.~J.}\ \bibnamefont
  {Liptak}} \emph {et~al.},\ }\href {\doibase 10.1016/j.nima.2022.167168}
  {\bibfield  {journal} {\bibinfo  {journal} {Nucl. Instrum. Meth. A}\ }\textbf
  {\bibinfo {volume} {1040}},\ \bibinfo {pages} {167168} (\bibinfo {year}
  {2022})}\BibitemShut {NoStop}%
\bibitem [{\citenamefont {Kuhr}\ \emph {et~al.}(2019)\citenamefont {Kuhr},
  \citenamefont {Pulvermacher}, \citenamefont {Ritter}, \citenamefont {Hauth},\
  and\ \citenamefont {Braun}}]{Kuhr:2018lps}%
  \BibitemOpen
  \bibfield  {author} {\bibinfo {author} {\bibfnamefont {T.}~\bibnamefont
  {Kuhr}}, \bibinfo {author} {\bibfnamefont {C.}~\bibnamefont {Pulvermacher}},
  \bibinfo {author} {\bibfnamefont {M.}~\bibnamefont {Ritter}}, \bibinfo
  {author} {\bibfnamefont {T.}~\bibnamefont {Hauth}}, \ and\ \bibinfo {author}
  {\bibfnamefont {N.}~\bibnamefont {Braun}} (\bibinfo {collaboration} {Belle II
  Framework Software Group}),\ }\href {\doibase 10.1007/s41781-018-0017-9}
  {\bibfield  {journal} {\bibinfo  {journal} {Comput. Software Big Sci.}\
  }\textbf {\bibinfo {volume} {3}},\ \bibinfo {pages} {1} (\bibinfo {year}
  {2019})}\BibitemShut {NoStop}%
\bibitem [{bas()}]{basf2-zenodo}%
  \BibitemOpen
  \href@noop {} {}\bibinfo {howpublished}
  {\url{https://doi.org/10.5281/zenodo.5574115}}\BibitemShut {NoStop}%
\bibitem [{\citenamefont {Bertacchi}\ \emph {et~al.}(2021)\citenamefont
  {Bertacchi} \emph {et~al.}}]{Bertacchi:2020eez}%
  \BibitemOpen
  \bibfield  {author} {\bibinfo {author} {\bibfnamefont {V.}~\bibnamefont
  {Bertacchi}} \emph {et~al.} (\bibinfo {collaboration} {Belle II Tracking
  Group}),\ }\href {\doibase 10.1016/j.cpc.2020.107610} {\bibfield  {journal}
  {\bibinfo  {journal} {Comput. Phys. Commun.}\ }\textbf {\bibinfo {volume}
  {259}},\ \bibinfo {pages} {107610} (\bibinfo {year} {2021})}\BibitemShut
  {NoStop}%
\bibitem [{\citenamefont {Milesi}\ \emph {et~al.}(2020)\citenamefont {Milesi},
  \citenamefont {Tan},\ and\ \citenamefont {Urquijo}}]{ebdt}%
  \BibitemOpen
  \bibfield  {author} {\bibinfo {author} {\bibfnamefont {M.}~\bibnamefont
  {Milesi}}, \bibinfo {author} {\bibfnamefont {J.}~\bibnamefont {Tan}}, \ and\
  \bibinfo {author} {\bibfnamefont {P.}~\bibnamefont {Urquijo}},\ }\href
  {\doibase 10.1051/epjconf/202024506023} {\bibfield  {journal} {\bibinfo
  {journal} {EPJ Web Conf.}\ }\textbf {\bibinfo {volume} {245}},\ \bibinfo
  {pages} {06023} (\bibinfo {year} {2020})}\BibitemShut {NoStop}%
\bibitem [{\citenamefont {Hulsbergen}(2005)}]{HULSBERGEN2005566}%
  \BibitemOpen
  \bibfield  {author} {\bibinfo {author} {\bibfnamefont {W.~D.}\ \bibnamefont
  {Hulsbergen}},\ }\href {\doibase https://doi.org/10.1016/j.nima.2005.06.078}
  {\bibfield  {journal} {\bibinfo  {journal} {Nucl. Instrum. Methods}\ }\textbf
  {\bibinfo {volume} {552}},\ \bibinfo {pages} {566} (\bibinfo {year}
  {2005})}\BibitemShut {NoStop}%
\bibitem [{\citenamefont {Krohn}\ \emph {et~al.}(2020)\citenamefont {Krohn}
  \emph {et~al.}}]{Krohn:2019dlq}%
  \BibitemOpen
  \bibfield  {author} {\bibinfo {author} {\bibfnamefont {J.-F.}\ \bibnamefont
  {Krohn}} \emph {et~al.} (\bibinfo {collaboration} {Belle II Analysis Software
  Group}),\ }\href {\doibase 10.1016/j.nima.2020.164269} {\bibfield  {journal}
  {\bibinfo  {journal} {Nucl. Instrum. Methods Phys. Res., Sect. A}\ }\textbf
  {\bibinfo {volume} {{976}}},\ \bibinfo {pages} {164269} (\bibinfo {year}
  {2020})}\BibitemShut {NoStop}%
\bibitem [{\citenamefont {Waltenberger}\ \emph {et~al.}(2008)\citenamefont
  {Waltenberger}, \citenamefont {Mitaroff}, \citenamefont {Moser},
  \citenamefont {Pflugfelder},\ and\ \citenamefont
  {Riedel}}]{Waltenberger_2008}%
  \BibitemOpen
  \bibfield  {author} {\bibinfo {author} {\bibfnamefont {W.}~\bibnamefont
  {Waltenberger}}, \bibinfo {author} {\bibfnamefont {W.}~\bibnamefont
  {Mitaroff}}, \bibinfo {author} {\bibfnamefont {F.}~\bibnamefont {Moser}},
  \bibinfo {author} {\bibfnamefont {B.}~\bibnamefont {Pflugfelder}}, \ and\
  \bibinfo {author} {\bibfnamefont {H.~V.}\ \bibnamefont {Riedel}},\ }\href
  {\doibase 10.1088/1742-6596/119/3/032037} {\bibfield  {journal} {\bibinfo
  {journal} {J. Phys. Conf. Ser.}\ }\textbf {\bibinfo {volume} {119}},\
  \bibinfo {pages} {032037} (\bibinfo {year} {2008})}\BibitemShut {NoStop}%
\bibitem [{\citenamefont {Dey}\ and\ \citenamefont
  {Soffer}(2020)}]{btube-conf}%
  \BibitemOpen
  \bibfield  {author} {\bibinfo {author} {\bibfnamefont {S.}~\bibnamefont
  {Dey}}\ and\ \bibinfo {author} {\bibfnamefont {A.}~\bibnamefont {Soffer}},\
  }\href@noop {} {\bibfield  {journal} {\bibinfo  {journal} {Springer Proc.
  Phys.}\ }\textbf {\bibinfo {volume} {248}},\ \bibinfo {pages} {411} (\bibinfo
  {year} {2020})}\BibitemShut {NoStop}%
\bibitem [{\citenamefont {{Ed.~A.~J.~Bevan, B.~Golob, Th.~Mannel, S.~Prell, and
  B.~D.~Yabsley}}(2014)}]{Bevan:2014iga}%
  \BibitemOpen
  \bibfield  {author} {\bibinfo {author} {\bibnamefont {{Ed.~A.~J.~Bevan,
  B.~Golob, Th.~Mannel, S.~Prell, and B.~D.~Yabsley}}},\ }\href {\doibase
  10.1140/epjc/s10052-014-3026-9} {\bibfield  {journal} {\bibinfo  {journal}
  {Eur. Phys. J.}\ }\textbf {\bibinfo {volume} {C74}},\ \bibinfo {pages} {3026}
  (\bibinfo {year} {2014})},\ \bibinfo {note} {see Section 6.5.2}\BibitemShut
  {NoStop}%
\bibitem [{\citenamefont {Fox}\ and\ \citenamefont
  {Wolfram}(1978)}]{PhysRevLett.41.1581}%
  \BibitemOpen
  \bibfield  {author} {\bibinfo {author} {\bibfnamefont {G.~C.}\ \bibnamefont
  {Fox}}\ and\ \bibinfo {author} {\bibfnamefont {S.}~\bibnamefont {Wolfram}},\
  }\href {\doibase 10.1103/PhysRevLett.41.1581} {\bibfield  {journal} {\bibinfo
   {journal} {Phys. Rev. Lett.}\ }\textbf {\bibinfo {volume} {41}},\ \bibinfo
  {pages} {1581} (\bibinfo {year} {1978})}\BibitemShut {NoStop}%
\bibitem [{\citenamefont {Chen}\ and\ \citenamefont
  {Guestrin}(2016)}]{ChenG16}%
  \BibitemOpen
  \bibfield  {author} {\bibinfo {author} {\bibfnamefont {T.}~\bibnamefont
  {Chen}}\ and\ \bibinfo {author} {\bibfnamefont {C.}~\bibnamefont
  {Guestrin}},\ }\bibfield  {booktitle} {\emph {\bibinfo {booktitle}
  {{Proceedings of the 22nd {ACM} {SIGKDD} International Conference on
  Knowledge Discovery and Data Mining}}},\ }\href@noop {} {\  (\bibinfo {year}
  {Association for Computing Machinery, New York, USA, 2016})}\BibitemShut
  {NoStop}%
\bibitem [{\citenamefont {Asner}\ \emph {et~al.}(1996)\citenamefont {Asner}
  \emph {et~al.}}]{PhysRevD.53.1039}%
  \BibitemOpen
  \bibfield  {author} {\bibinfo {author} {\bibfnamefont {D.~M.}\ \bibnamefont
  {Asner}} \emph {et~al.},\ }\href {\doibase 10.1103/PhysRevD.53.1039}
  {\bibfield  {journal} {\bibinfo  {journal} {Phys. Rev. D}\ }\textbf {\bibinfo
  {volume} {53}},\ \bibinfo {pages} {1039} (\bibinfo {year}
  {1996})}\BibitemShut {NoStop}%
\bibitem [{\citenamefont {Lee}\ \emph {et~al.}(2003)\citenamefont {Lee} \emph
  {et~al.}}]{PhysRevLett.91.261801}%
  \BibitemOpen
  \bibfield  {author} {\bibinfo {author} {\bibfnamefont {S.~H.}\ \bibnamefont
  {Lee}} \emph {et~al.} (\bibinfo {collaboration} {Belle Collaboration}),\
  }\href {\doibase 10.1103/PhysRevLett.91.261801} {\bibfield  {journal}
  {\bibinfo  {journal} {Phys. Rev. Lett.}\ }\textbf {\bibinfo {volume} {91}},\
  \bibinfo {pages} {261801} (\bibinfo {year} {2003})}\BibitemShut {NoStop}%
\bibitem [{\citenamefont {Bjorken}\ and\ \citenamefont
  {Brodsky}(1970)}]{PhysRevD.1.1416}%
  \BibitemOpen
  \bibfield  {author} {\bibinfo {author} {\bibfnamefont {J.~D.}\ \bibnamefont
  {Bjorken}}\ and\ \bibinfo {author} {\bibfnamefont {S.~J.}\ \bibnamefont
  {Brodsky}},\ }\href {\doibase 10.1103/PhysRevD.1.1416} {\bibfield  {journal}
  {\bibinfo  {journal} {Phys. Rev. D}\ }\textbf {\bibinfo {volume} {1}},\
  \bibinfo {pages} {1416} (\bibinfo {year} {1970})}\BibitemShut {NoStop}%
\bibitem [{\citenamefont {Brandt}\ \emph {et~al.}(1964)\citenamefont {Brandt},
  \citenamefont {Peyrou}, \citenamefont {Sosnowski},\ and\ \citenamefont
  {Wroblewski}}]{Brandt:1964sa}%
  \BibitemOpen
  \bibfield  {author} {\bibinfo {author} {\bibfnamefont {S.}~\bibnamefont
  {Brandt}}, \bibinfo {author} {\bibfnamefont {C.}~\bibnamefont {Peyrou}},
  \bibinfo {author} {\bibfnamefont {R.}~\bibnamefont {Sosnowski}}, \ and\
  \bibinfo {author} {\bibfnamefont {A.}~\bibnamefont {Wroblewski}},\ }\href
  {\doibase 10.1016/0031-9163(64)91176-X} {\bibfield  {journal} {\bibinfo
  {journal} {Phys. Lett.}\ }\textbf {\bibinfo {volume} {12}},\ \bibinfo {pages}
  {57} (\bibinfo {year} {1964})}\BibitemShut {NoStop}%
\bibitem [{\citenamefont {Farhi}(1977)}]{Farhi:1977sg}%
  \BibitemOpen
  \bibfield  {author} {\bibinfo {author} {\bibfnamefont {E.}~\bibnamefont
  {Farhi}},\ }\href {\doibase 10.1103/PhysRevLett.39.1587} {\bibfield
  {journal} {\bibinfo  {journal} {Phys. Rev. Lett.}\ }\textbf {\bibinfo
  {volume} {39}},\ \bibinfo {pages} {1587} (\bibinfo {year}
  {1977})}\BibitemShut {NoStop}%
\bibitem [{\citenamefont {Cheema}(2024)}]{pbdt}%
  \BibitemOpen
  \bibfield  {author} {\bibinfo {author} {\bibfnamefont {P.}~\bibnamefont
  {Cheema}},\ }\href {\doibase 10.1051/epjconf/202429509035} {\bibfield
  {journal} {\bibinfo  {journal} {EPJ Web of Conf.}\ }\textbf {\bibinfo
  {volume} {295}},\ \bibinfo {pages} {09035} (\bibinfo {year}
  {2024})}\BibitemShut {NoStop}%
\bibitem [{\citenamefont {Gaiser}(1982)}]{Gaiser:Phd}%
  \BibitemOpen
  \bibfield  {author} {\bibinfo {author} {\bibfnamefont {J.}~\bibnamefont
  {Gaiser}},\ }\href@noop {} {Ph.D. thesis},\ \bibinfo  {school} {Stanford
  University} (\bibinfo {year} {1982})\BibitemShut {NoStop}%
\bibitem [{\citenamefont {Skwarnicki}(1986)}]{Skwarnicki:1986xj}%
  \BibitemOpen
  \bibfield  {author} {\bibinfo {author} {\bibfnamefont {T.}~\bibnamefont
  {Skwarnicki}},\ }\href@noop {} {Ph.D. thesis},\ \bibinfo  {school} {Cracow,
  INP} (\bibinfo {year} {1986})\BibitemShut {NoStop}%
\bibitem [{\citenamefont {Banerjee}\ \emph {et~al.}()\citenamefont {Banerjee}
  \emph {et~al.}}]{Banerjee:2024znd}%
  \BibitemOpen
  \bibfield  {author} {\bibinfo {author} {\bibfnamefont {S.}~\bibnamefont
  {Banerjee}} \emph {et~al.} (\bibinfo {collaboration} {{HFLAV
  Collaboration}}),\ }\href@noop {} {\ }\Eprint
  {http://arxiv.org/abs/2411.18639} {arXiv:2411.18639} \BibitemShut {NoStop}%
\bibitem [{\citenamefont {Pivk}\ and\ \citenamefont
  {Le~Diberder}(2005)}]{Pivk:2004ty}%
  \BibitemOpen
  \bibfield  {author} {\bibinfo {author} {\bibfnamefont {M.}~\bibnamefont
  {Pivk}}\ and\ \bibinfo {author} {\bibfnamefont {F.~R.}\ \bibnamefont
  {Le~Diberder}},\ }\href {\doibase 10.1016/j.nima.2005.08.106} {\bibfield
  {journal} {\bibinfo  {journal} {Nucl. Instrum. Methods Phys. Res., Sect. A}\
  }\textbf {\bibinfo {volume} {{555}}},\ \bibinfo {pages} {356} (\bibinfo
  {year} {2005})}\BibitemShut {NoStop}%
\bibitem [{\citenamefont {Long}\ \emph {et~al.}(2003)\citenamefont {Long},
  \citenamefont {Baak}, \citenamefont {Cahn},\ and\ \citenamefont
  {Kirkby}}]{PhysRevD.68.034010}%
  \BibitemOpen
  \bibfield  {author} {\bibinfo {author} {\bibfnamefont {O.}~\bibnamefont
  {Long}}, \bibinfo {author} {\bibfnamefont {M.}~\bibnamefont {Baak}}, \bibinfo
  {author} {\bibfnamefont {R.~N.}\ \bibnamefont {Cahn}}, \ and\ \bibinfo
  {author} {\bibfnamefont {D.}~\bibnamefont {Kirkby}},\ }\href {\doibase
  10.1103/PhysRevD.68.034010} {\bibfield  {journal} {\bibinfo  {journal} {Phys.
  Rev. D}\ }\textbf {\bibinfo {volume} {68}},\ \bibinfo {pages} {034010}
  (\bibinfo {year} {2003})}\BibitemShut {NoStop}%
\end{thebibliography}%


\end{document}